\documentclass[11pt]{article}
\usepackage{graphicx}
\usepackage[authoryear,round]{natbib}

\makeatletter

\usepackage{amsmath,multirow}
\usepackage{enumerate,array}
\usepackage{amsfonts}
\usepackage{mathtools}
\usepackage{amssymb}
\usepackage{epsfig}
\usepackage{graphics}
\usepackage{multicol, multirow}
\usepackage[pdfstartview=XYZ, bookmarks=true, colorlinks=true, linkcolor=blue, urlcolor=blue, citecolor=blue, pdftex, bookmarks=true, linktocpage=true, hyperindex=true]{hyperref}
\usepackage{enumitem}
\usepackage{color}
\usepackage{amsthm}
\usepackage{float}
\usepackage{lscape}
\usepackage{verbatim}
\usepackage{amsfonts}
\usepackage{lineno}
\usepackage{xr-hyper}
\usepackage[nameinlink]{cleveref}

\newcommand{\beq}{\begin{equation}}
\newcommand{\eeq}{\end{equation}}

\makeatletter
\newcommand{\pushright}[1]{\ifmeasuring@#1\else\omit\hfill$\displaystyle#1$\fi\ignorespaces}
\newcommand{\pushleft}[1]{\ifmeasuring@#1\else\omit$\displaystyle#1$\hfill\fi\ignorespaces}
\makeatother

\renewcommand{\baselinestretch}{1.2}

\begin{document}
	
	\title{Generalized method of L-moment estimation \\ for stationary and nonstationary extreme value models}
	
	\author{Yonggwan Shin$^{1}$,  ~Yire Shin$^{2, *}$, ~Jihong Park$^{3}$, ~ Jeong-Soo Park$^{2,4}$  \\	\\
		\small \it 1: R\&D Center, XRAI Inc., Gwangju 61186, Korea \\ 
			\small \it 2: Department of Statistics, Chonnam National University, Gwangju 61186, Korea \\
   \small \it 3: Department of Mathematics and Statistics, \\ 
   	\small \it Chonnam National University, Gwangju 61186, Korea \\
		\small \it 4: Research Unit of Data Science for Sustainable Agriculture, \\ \small \it Mahasarakham University, Maha Sarakham 44150, Thailand \\
		\small \it *: Corresponding author, e-mail: shinyire87@gmail.com 
	}
	\maketitle  

\begin{abstract}
	
	 Precisely estimating out-of-sample upper quantiles is very important in risk assessment and in engineering practice for structural design to prevent a greater disaster. For this purpose, the generalized extreme value (GEV) distribution has been broadly used. To estimate the parameters of GEV distribution, the maximum likelihood estimation (MLE) and L-moment estimation (LME) methods have been primarily employed.
	 For a better estimation using the MLE, several studies considered the generalized MLE (penalized likelihood or Bayesian) methods to cooperate with a penalty function or prior information for parameters. However, a generalized LME method for the same purpose has not been developed yet in the literature. 
We thus propose the generalized method of L-moment estimation (GLME) to cooperate with a penalty function or prior information. The proposed estimation is based on the generalized L-moment distance and a multivariate normal likelihood approximation. Because the L-moment estimator is more efficient and robust for small samples than the MLE, we reasonably expect the advantages of LME to continue to hold for GLME. The proposed method is applied to the stationary and nonstationary GEV models with two novel (data-adaptive) penalty functions to correct the bias of LME. A simulation study indicates that the biases of LME are considerably corrected by the GLME with slight increases in the standard error. Applications to US flood damage data and maximum rainfall at Phliu Agromet in Thailand illustrate the usefulness of the proposed method. This study may promote further work on penalized or Bayesian inferences based on L-moments. 
\end{abstract}

\vspace{5mm}
\begin{itemize}
	\item {\bf Research highlight 1}: The generalized method of L-moment estimation to cooperate with a prior or a penalty function is proposed.
	\item {\bf Research highlight 2}: The proposed method is applied to both the stationary and nonstationary extreme value models.
	\item {\bf Research highlight 3}: Bias correction for the L-moment estimator is attempted with two (data-adaptive) penalty functions to better estimate the return levels for long return periods.	
	\item {\bf Research highlight 4}: Applications to two extreme hydrological data illustrate the usefulness of proposed method.
\end{itemize}

\vspace{5mm} \noindent {\bf Keywords}: Bias correction; Data-adaptive penalty function;  Generalized L-moment distance; Rainfall extremes; Return level.

\subsection*{Abbreviation list ----------------------------------------------------------------}
\begin{itemize}
\item {\bf $B(p,q)$}: Beta function with inputs $p$ and $q$
\item {\bf CD PF}: Coles-Dixon Penalty function
\item {\bf GEV}: Generalized extreme value
\item {\bf GEVD}: GEV  distribution
\item {\bf GEV11 model}: Time-dependent nonstationary GEV model with location parameter $\mu_t = \mu_0 + \mu_1 \times t\ $ and scale parameter $ \ \sigma_t = \text{exp}\ (\sigma_0 + \sigma_1 \times t)$.
\item {\bf GMLE}: Generalized maximum likelihood estimator
\item {\bf GLD or $\omega(\mu,\sigma,\xi)$}: Generalized L-moment distance
\item {\bf GLME}: Generalized L-moment estimator
\item {\bf GLME.b}: GLME with a beta penalty function 
\item {\bf GLME.n}: GLME with a normal penalty function 
\item {\bf LME}: L-moment estimator 
\item {\bf MLE}: Maximum likelihood estimator 
\item {\bf MS PF}: Martins-Stedinger Penalty function
\item {\bf NS}: Nonstationary
\item {\bf PDF}: Probability density function
\item {\bf PF}: Penalty function
\item {\bf RMSE}: Root mean squared error 
\item {\bf SE}: Standard error 
\item {\bf $\mu,\, \sigma,\, \xi$}: Location, scale, and shape parameters of GEVD
\item {\bf $p(\xi)$}: Penalty function on the shape parameter of GEVD
\\
---------------------------------------------------------------------------------------------
\end{itemize}

\section{Introduction}

Precise estimation of quantities such as so-called T-year return levels of extreme values is an important step in engineering practice for structural design to prevent a greater disaster. The generalized extreme value (GEV) distribution has been broadly employed for this purpose, when the block maxima data are given. In estimating the parameters of GEV distribution, the L-moment estimation (LME) and maximum likelihood estimation (MLE) methods have been primarily used. \citep{coles2001introduction, hosking1997regional}. However, estimating out-of-sample upper quantiles with a small sample size is often unsatisfactory, especially when larger uncertainties are involved.
 
For better MLE performance, some authors have suggested using the penalized or generalized MLE method, which cooperates with a penalty function (PF) or a prior information for parameters of extreme value models. \cite{coles1999likelihood} proposed a penalized likelihood method using a PF on the shape parameter of the GEV model. \cite{martins2000generalized, martins2001generalized} improved the MLE using a PF (or a prior) based on a beta probability density function. They referred to the method as the generalized MLE (GMLE). Several authors \citep{cannon2010flexible, bucher2021penalized, papukdee2022penalized, castro2022practical} considered other kinds of PFs or penalized likelihood methods. The GMLE works well when the given prior is nearly correct, but performs poorly when the given PF is far from the ground truth \citep{park2005simulation, lee2017data, wehner2024uncertainty}.

 For small or moderate samples, it is known that the LME is usually more efficient, computationally simpler, and more robust than the MLE \citep{karvanen2006estimation, delicado2008small, asquith2014parameter, lillo2016moments, nerantzaki2022assessing, wehner2024uncertainty}. 
While the GMLE method is available readily to estimate GEV parameters, the L-moments method is not, because applying the LME with a PF for parameters is not straightforward; therefore, a method for doing so for the L-moments has not been reported yet, to the best of our knowledge. \cite{wehner2024uncertainty} stated that ``despite the greater efficiency of the L-moments method, the MLE methods would generally be preferable, because there are no such practical implementations for L-moments" that deal with nonstationary data (or deal with prior information of parameters).
Thus, this study proposes and implements a new method for the LME to cooperate with prior information for parameters, which is applicable to both stationary and nonstationary extreme value models. The proposed method is referred to as the generalized method of L-moments estimation (GLME), which is analogous to the (Bayesian) generalized method of moment estimation \citep{hansen1982large, zellner1997bayesian, Yin2009BayesianGM}. The advantages of the LME are reasonably expected to remain valid when estimating the parameters by the GLME with a suitably chosen PF or a prior.

A few studies considered using Bayesian inference and the L-moments together. For example, \cite{dodangeh2014application} applied the L-moments for estimating low-flow quantiles and carried out uncertainty analysis using Bayesian inference. \cite{moges2018bayesian} employed Bayesian model averaging approach augmented with L-moments for regional frequency analysis. \cite{jayaraman2023moments} used the sample L-moments for Bayesian approach for probabilistic risk assessment and uncertainty quantification. These studies applied L-moments and Bayesian methods together, but did not tried the Bayesian (or generalized) LME for parameters of extreme value models.

This study proposes a GLME method with two new penalty functions and demonstrates its usefulness for analyzing block maxima data. Section 2 describes the LME and GMLE. Next, Sections 3 and 5 detail the proposed GLME method applied to stationary and nonstationary GEV models, while Section 4 describes the penalty functions. Then, Section~6 offers a simulation study comparing the performance of GLME method with that of LME. Section 7 illustrates the utility of the GLME by applying it to two real datasets. Moreover, Sections 8 and 9 present the discussion and conclusion. The Supplementary Material provides details, including tables and figures. The first version of the R code and the data for this study are available at GitHub depository, https://github.com/sygstat/GL-momentEst.git.

\section{Extreme value model and estimation}
\subsection{Generalized extreme value distribution}
The GEV distribution (GEVD) has been broadly applied to model the extreme events of human society and natural phenomena \citep{naghettini2017fundamentals, bousquet2021extreme}. The main reason why the GEVD has been widely used is that it is theoretically derived as the size of block goes to infinity \citep{coles2001introduction, embrechts2013modelling}. The stationary GEVD has the following probability density function (PDF)  \citep{hosking1997regional, tawn1988extreme}:
\begin{equation} \label{pdf-gevd}
f(x)\ =\ \sigma^{-1} \left(1 -\xi {{x-\mu} \over {\sigma}}\right) ^{{1 \over \xi} -1}\times 
\exp \left\{ - \left(1 -\xi {{x-\mu} \over {\sigma}}\right) ^{1/ \xi} \right\}, 
\end{equation}
 under the condition that $1-\xi (x- \mu ) / \sigma > 0$ and $\sigma>0$. In this formula, $\mu, \; \sigma$, and $\xi$ are the location, scale, and shape parameters, respectively. We have to note that the sign of $\xi$ is opposite to the formula in the book by \cite{coles2001introduction}. The GEVD has various tails according to the sign of the shape parameter $\xi$: a heavy tail when $\xi<0$, exponential when $\xi \rightarrow 0$, and a light tail when $\xi >0$.

The return level $r_T$ is defined as the $1-p$ quantile of the distribution associated with the return period $T=1/p$ \citep{coles2001introduction}. The name `T-year return level’ is employed for the annual extreme data. Researchers usually try to accurately estimate the return levels for long return periods such as $T=50$ at least or $T=100$. 

\subsection{L-moment  estimation}
\cite{hosking1990moments} established L-moments as a linear combination of the expectations of order statistics. The $r$th population L-moments ($\lambda_r$) of a probability distribution are specified as follows: 
\begin{eqnarray} \label{l-moments}
\lambda_1& =& E (X_{1:1} ), \\
\lambda_2& =& {1\over2}\ E (X_{2:2}- X_{1:2}), \\
\lambda_3& =& {1\over3}\ E(X_{3:3}-2 X_{2:3}+ X_{1:3}), 
\end{eqnarray}
and in general
\begin{equation}
\lambda_r\ =\ r^{-1}\ \sum_{j=0}^{r-1}\ (-1)^j\ {{r-1}\choose{j}}\ E(X_{r-j:r}), 
\end{equation}
where $X_{k:n}$ denotes the $k$th smallest order statistic from a sample of size of $n$. 

The $r$th sample L-moments ($l_r$) are obtained as unbiased estimates of $\lambda_r$ as follows \citep{wang1996direct}:
\begin{equation}
	\begin{aligned}
	& l_1 = \frac{1}{n} \sum x_i , ~~~~ l_2 = \frac {1}{2} {n\choose 2}^{-1}
	\sum\sum_{j<i} (x_{i:n} - x_{j:n} ) , \\
	& l_3 = \frac{1}{3} {n\choose
		3}^{-1} \sum\sum\sum_{k< j < i} (x_{i:n} - 2x_{j:n} + x_{k:n}
	), {\nonumber}\\ \vdots {\nonumber} \\
	& l_r = \frac{1}{r} {n\choose r}^{-1} \sum_{ i_1} \sum_{\leq\, i_2} \dots
	\sum_{\leq\, i_n}\,  \sum_{k=0}^{r-1} (-1)^k {r-1 \choose k} x_{i_{r-k} :n} ,\; ~ r=1,2,\dots, n.
	\end{aligned}
\end{equation}
Thus $l_r$ is a linear combination of the sample order statistics $x_{1:n},\dots, x_{n:n}$. The L-moments provide theoretical and practical superiorities over ordinary moments. For instance, L-moments are less affected by outliers and better identify the parent distribution that produces a sample \citep{hosking1997regional, silva2020moments}. 

Analogous to the typical method of moment estimation, the method of L-moments estimation acquires parameter estimates by equating the first $r$ population L-moments to the corresponding sample quantities. The LME is often robust to outliers, more efficient, and computationally easier than the MLE, for small and moderate samples. Thus, the LME has been extensively applied in diverse research areas, including hydrology, atmospheric science, and engineering \citep[for example]{karvanen2006estimation, delicado2008small, asquith2014parameter, murshed2014lh, lillo2016moments, simkova2021confidence, yilmaz2021comparison, hong2022changes}. However, one problem of the LME is that it sometimes underestimates high quantiles when the shape parameter of the GEVD is less than $-0.2$. Aiming to address this problem, the following Section 4 considers two kinds of PFs.

\subsection{Generalized maximum likelihood estimation}
Assuming that the observations $x_1, x_2,\dots, x_n$ are sampled from the GEVD, the negative log-likelihood function of $(\mu, \sigma, \xi)$ is
\begin{equation}\label{gevnllh}
l(\mu, \sigma, \xi)\ =\ n\ln\sigma\ +\ (1-1/\xi)\sum_{i=1}^{n}
\ln\ \{1-\xi({x_i}-\mu)/\sigma \} \ +\ \sum_{i=1}^{n}\ \{1-\xi({x_i}-\mu)/\sigma\}^{1/\xi}, 
\end{equation}
when $1-\xi({{x_i}-\mu})/{\sigma} >0 \mbox{ for } i=1,\dots, n$. Then 
the MLE of $(\mu, \sigma, \xi)$ are obtained from the minimum of Eq.~(\ref{gevnllh}). The R package `ismev' was employed in this study to obtain the MLE \citep{coles2001introduction}.

The MLE sometimes performs poorly for a small sample and underestimates the negative value of $\xi$, so that overestimates return values, especially for $\xi < -.25$. Consequently, the MLE may cause a large variance and bias for high upper quantiles for such cases. To overcome this trouble, \cite{coles1999likelihood} 
and \cite{martins2000generalized, martins2001generalized} 
proposed the GMLE methods with PFs on $\xi$.
\cite{el2007generalized} extended the GMLE method to nonstationary GEV model using Bayesian computation.

 The GMLE is gained by minimizing the following generalized (or penalized) negative log-likelihood function with respect to $(\mu, \sigma, \xi)$:
\begin{equation} \label{GMLE}
l_{G}\ (\mu, \sigma, \xi)\ =\ l(\mu, \sigma, \xi) \ - \ \ln ( p(\xi)), 
\end{equation}
where $p(\xi)$ denotes a PF on $\xi$.

\section{Generalized method of L-moment estimation (GLME)}
\subsection{Generalized L-moment distance}
The system of equations to obtain the LME are:
\begin{eqnarray} \label{lambda-l}
\underline {\bf \lambda}\ - \ \underline{\bf l}\ = \ \underline{\bf 0}, 
\end{eqnarray}
where $\underline{\bf \lambda} =(\lambda_1, \; \lambda_2, \; \lambda_3)^t$ and 
$\underline{\bf l} =(l_1, \; l_2, \; l_3)^t$. From these equations, the generalized L-moment distance (GLD) is defined as follows:
\begin{eqnarray} \label{GLMDist}
GLD(\mu, \sigma, \xi)\ =\ \omega(\mu, \sigma, \xi)\ = \ (\underline{\bf \lambda} -\underline{\bf l})^t\ V^{-1}\ (\underline{\bf \lambda} -\underline{\bf l}) , 
\end{eqnarray}
where $V$ represents the covariance matrix of the sample L-moments up to the third order. The matrix $V$ is computed from the expression by \cite{elamir2004exact}. This computation was performed using the `lmomco' package \citep{asquith2011distributional} in R software. In rare cases, we experienced problems in inverting $V$ because it was not nonsingular numerically. For these cases, $V$ was constructed using a nonparametric bootstrap method. Versions of GLD have been used by some researchers \citep{ kjeldsen2015bivariate, alvarez2023, ShinPark2024Serra, shin2025modelaveraging_arXiv}. 

\subsection{GLME using the multivariate normal likelihood}
Because the LME satisfying Eq.~(\ref{lambda-l}) implies that the GLD is zero, the LME is also obtained by minimizing the following function:
\begin{eqnarray} \label{MNorm}
f\ (\omega(\mu, \sigma, \xi))\ =\ \frac{1}{(2 \pi)^{3/2}\ |V| }\;
\exp \{-{1\over2}\ \omega(\mu, \sigma, \xi) \}.
\end{eqnarray}
This function is a PDF of the three-variate normal random vector. As sample L-moments typically converge to a multivariate normal distribution as $n \rightarrow \infty$ \citep{hosking1990moments}, this function can be treated as an approximation to the likelihood function of $\underline{\bf \theta}=(\mu, \sigma, \xi)$ given three sample L-moments. 

Then, the GLME of $\underline{\bf \theta}=(\mu, \sigma, \xi)$ is obtained by minimizing the following function with respect to $\underline{\bf \theta}$:
\begin{eqnarray} \label{GLME_func}
-\ln\ ( f\ (\omega(\underline{\bf \theta})))\; -\ \alpha_n \; \ln ( p(\underline{\bf \theta})), 
\end{eqnarray}
where $\alpha_n$ denotes the weight for the PF $p(\underline{\bf \theta})$ compared to the approximated likelihood of $\omega(\underline{\bf \theta})$. 
This study sets $\alpha_n=1$, but $\alpha_n$ can be selected from the data using the technique such as the cross-validation procedure minimizing the sum of empirical cross-entropies which was employed in \cite{bucher2021penalized}.

Notably, when $p(\underline{\bf \theta})$ is a prior probability function and $\alpha_n=1$, Eq.~(\ref{GLME_func}) is a negative log posterior probability function in the Bayesian paradigm. Minimizing (\ref{GLME_func}) with respect to $\underline{\bf \theta}$ is the same as obtaining the maximum a posteriori estimate of parameters in Bayesian statistics. See \cite{coles2005bayesian, coles1996bayesian} for a good application and a review of Bayesian methods in extreme value modeling. The proposed method is analogous to the Bayesian generalized method of moment estimation by \cite{Yin2009BayesianGM}. Instead of using the name Bayesian LME, this study uses the name `generalized' LME for consistency with the `generalized' MLE \citep{martins2000generalized, martins2001generalized, el2007generalized, hundecha2008nonstationary, cannon2010flexible, gilleland2016extremes, Zhang_GMLE_10363225}. 

\section{Penalty functions for bias correction in GEV model}

\subsection{Existing penalty functions}

The MLE of the GEV distribution often underestimates the shape parameter ($\xi$), especially for $\xi \le -0.2$ and for small samples. Consequently, it leads to a large positive bias in very high quantiles. To address this problem, \cite{coles1999likelihood} suggested a likelihood-based penalized inference, which allocates a penalty for a large negative value of $\xi$.

\cite{coles1999likelihood} proposed a exponential PF in the GEVD:
\begin{equation}\label{p_coles}
p(\xi) = \begin{cases} 1 & \mbox{if $\xi \geq 0$}\\
\exp\{-\lambda(\frac{1}{1+\xi}-1)^{\alpha}\} & \mbox{if $-1< \xi < 0$}\\
0 & \mbox{if $\xi\leq -1$}\end{cases}
\end{equation}
where $\alpha$ and $\lambda$ are non-negative hyperparameters. They suggested using $\alpha=1$ and $\lambda=1$. We call this as Coles-Dixon (CD) PF.

\cite{martins2000generalized} considered the generalized MLE approach by proposing a PF (or a prior) on $\xi$. The PF is based on the PDF of a beta distribution between $-0.5$ and $0.5$:
\begin{eqnarray} \label{MSpenalty}
p(\xi)= \frac{(0.5+\xi)^{p-1} (0.5-\xi)^{q-1}}{B(p,q)},
\end{eqnarray}
where $B(p,q)$ is the beta function on $(-0.5,\, 0.5)$. Their choice for $p$ and $q$ based on their experiments and prior hydrological information are $p=6$ and $q=9$. We abbreviate this function as Martins-Stedinger (MS) PF. Meanwhile, \cite{park2005simulation} and \cite{cannon2010flexible} suggested using $p=2.5,\, q=2.5$, and $p=2,\, q=3.3$, respectively. 
\cite{papukdee2022penalized} considered the modified versions of these PFs which were adjusted for an extended interval such as $(-1,\, 0.5)$. 

All of these PFs were considered to assign penalty to a large negative value of $\xi$ so as to reduce a positive bias of high upper quantiles by the MLE. These PFs perform a role to prevent the estimate of $\xi$ being a large negative value. Whereas, for applying a PF to correct the bias of LME, we have an opposite situation to the above PFs. That is, we need the PF which can promote the estimate of $\xi$ being a larger negative value than the LME of $\xi$. Thus, we may not employ the CD (or MS) PFs straightforwardly for the GLME. We rather propose other PFs in the next sections, especially for the bias correction of LME.


\subsection{Fixed normal penalty}

For the GLME, this study employs two PFs on the shape parameter. The first one is a ``fixed" PF, similarly fixed as the CD and MS PFs, which does not depend on the data. Whereas, the second one is the ``data-adaptive" PF, similarly as \cite{lee2017data}, which depends on the parameter estimates obtained from data.  

\begin{figure}[!htb]
	\centering 
	\includegraphics[width=11cm, height=8cm]{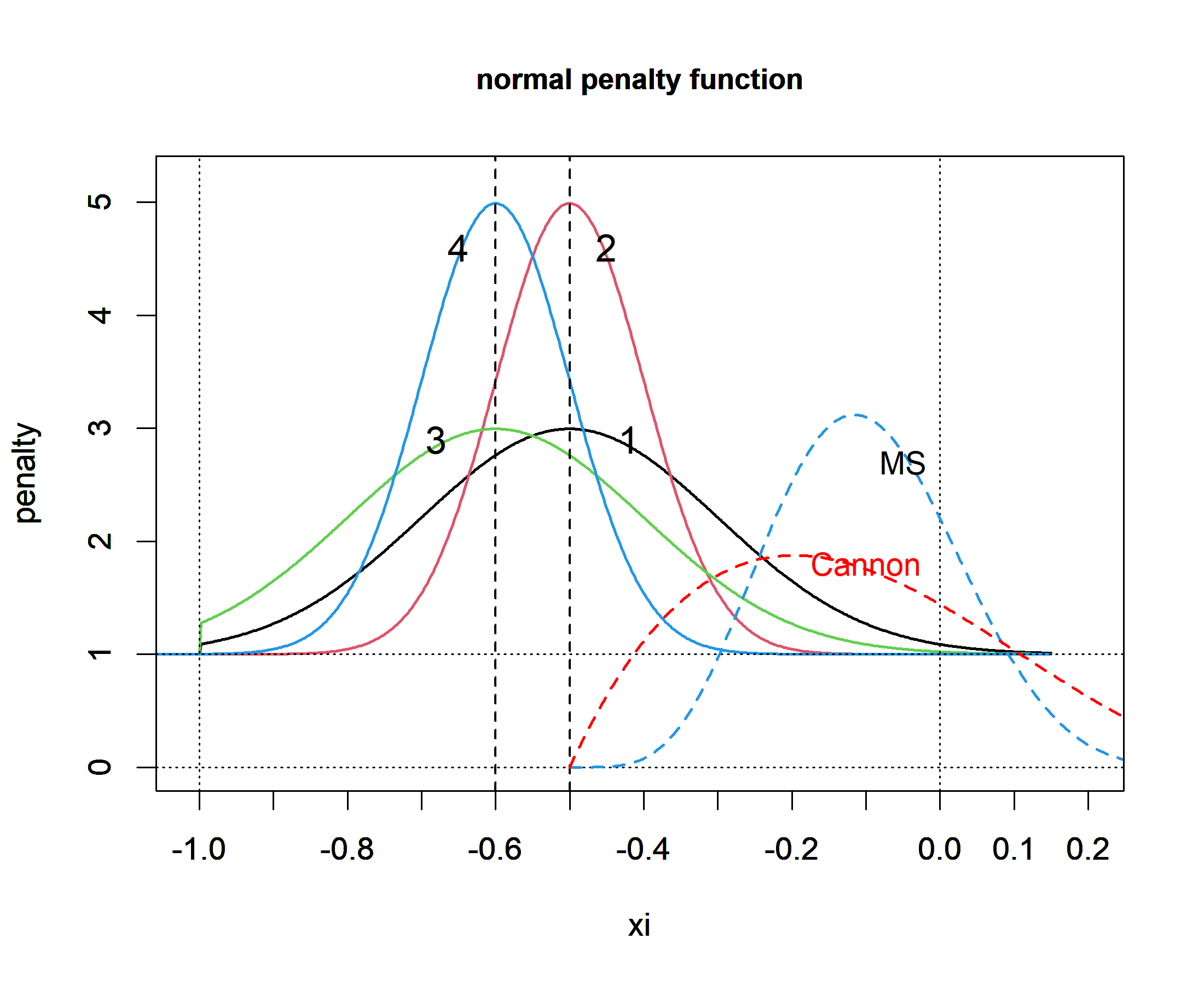} 
	\caption{Graphs of normal penalty functions (PFs) for choices of hyperparameters, Coles-Dixon (CD), and Cannon's PFs. The number on the graph of normal PF specifies the choice-number in (\ref{choice_hyper.n}).}
	\label{fig:pen.norm}
\end{figure}

The first one is a PF with a normal density:
\begin{eqnarray} \label{pen.norm}
p(\xi)\ = 1 \, +\, \psi (\xi;\: \mu_\xi, \sigma_\xi)
\end{eqnarray}
where $\psi (\xi;\: \mu_\xi, \sigma_\xi)$ represents a normal PDF of $\xi$ with a mean $\mu_\xi$ and standard deviation $\sigma_\xi$. This PF is referred to as a `normal PF'. The behaviour of this PF depends on the choice of hyperparameters, $\mu_\xi$ and $\sigma_\xi$.  We consider the following four choices:
\begin{equation} \label{choice_hyper.n}
\begin{aligned}
&\text{choice-1:}~ \mu_\xi=-0.5,\, \sigma_\xi=0.2 \\
&\text{choice-2:}~ \mu_\xi=-0.5,\, \sigma_\xi=0.1\\
&\text{choice-3:}~ \mu_\xi=-0.6,\, \sigma_\xi=0.2 \\
&\text{choice-4:}~  \mu_\xi=-0.6,\, \sigma_\xi=0.1\\
\end{aligned}
\end{equation} 
Of course, one can choose other hyperparameters.
Figure~\ref{fig:pen.norm} shows graphs of these PFs for choices of hyperparameters, and CD and Cannon's PFs for comparison.

Because the normal PF consists of a normal PDF plus one, it does not actually give any penalty to $\xi$ par away from $\mu_\xi$. Instead, it only attracts $\xi$ weakly toward $\mu_\xi$. Thus, the effect of using this PF may be relatively weak.

\subsection{Data-adaptive beta penalty}
The second PF for bias correction of the LME is based on the beta PDF and the LME calculated already from the data:
\begin{eqnarray} \label{pen.beta}
p(\xi)\ =\ \frac{(-L+\xi)^{p-1}\ (U-\xi)^{q-1}}{ B(p, q)}
\end{eqnarray}
for $\xi \in (L, \ U)$, where $B(p, q)$ is the beta function defined on the interval $(L, \ U)$ with parameters $p$ and $q$. This PF is referred to as a `beta PF'.  We set $L =\text{max}(-1, \ {\hat \xi}-c_0)$ and $U = \text{min}(0.3, \ {\hat \xi}+c_0)$, where ${\hat \xi}$ is the LME of $\xi$. The interval $(L, \ U)$ is considered for the range of $\xi$ and for the support of the beta function. Actually, the interval $(L, \ U)$ is $({\hat \xi}-c_0,\, {\hat \xi}+c_0)$ in $(-1,\, 0.3)$. We set $c_0=0.3$ in this study, but one can change it to 0.2 or 0.4 or 0.5.

\begin{figure}[!htb]
	\centering 
	\includegraphics[width=15cm, height=12cm]{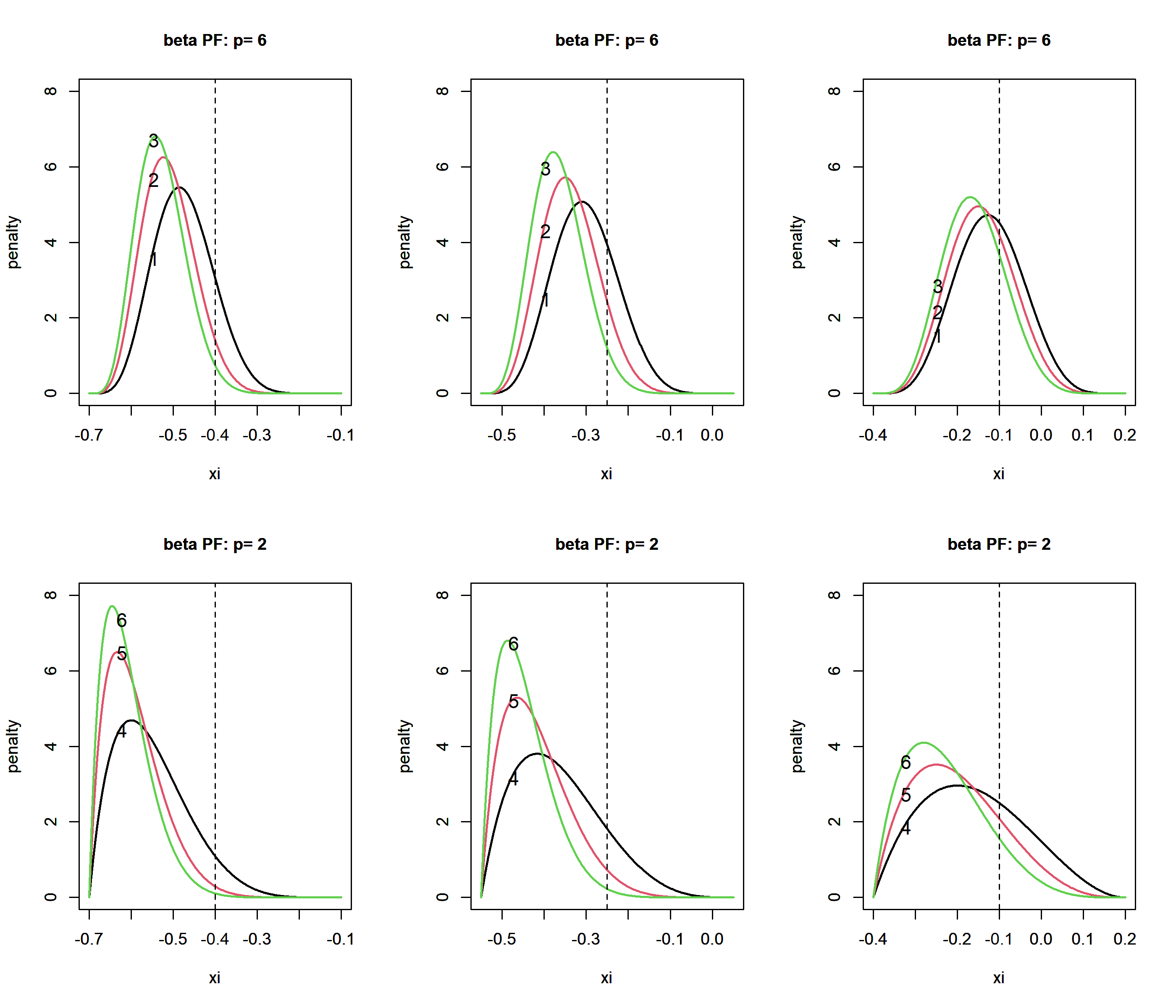} 
	\caption{Graphs of beta penalty functions (PFs) for choices of hyperparameters. Upper (lower) panel is for $p=6\, (2)$. The number on the graph of PF specifies the choice-number in Table~ \ref{tab:c1_c2}. The vertical dashed line represents the LME obtained from data which are $-0.4,\, -0.25$, and $-0.1$, respectively.}
	\label{fig:pen.beta}
\end{figure}

The behaviour of this PF depends on $\hat \xi$ and the choice of $p,\,q$. 
Our first choice for $p,\,q$ is $p=6$ and $q=p+a$, following \cite{martins2000generalized}, where $a= \text{min}(|{\hat \xi}| \times c_1, \ c_2)$ for ${\hat \xi} \le 0$ and $a=0$ for ${\hat \xi} >0$. This choice depends on $c_1,\, c_2$. For the values of $c_1,\, c_2$, we consider three sets: $(c_1=10,\, c_2=5)$, $(c_1=20,\, c_2=7)$, and $(c_1=30,\, c_2=9)$. As $c_1$ and $c_2$ increase, more preference for a larger negative value of $\xi$ is assigned, as seen in Figure~\ref{fig:pen.beta}, so that to lead to higher return level estimates than those from $\hat \xi$ (LME). The second choice for $p,\,q$ is $p=2$ and $q=p+a$, following \cite{cannon2010flexible}, where $a$ is the same as the above. Hence, we have six different sets of hyperparameters as provided in Table~\ref{tab:c1_c2}. The values of $q$ in Table ~\ref{tab:c1_c2} are obtained for three fixed estimates of $\xi$ that $\hat \xi = -0.4,\, -0.25,\, -0.1$, for example.

\begin{table}[htb]
	\caption{Choice of hyperparameters ($c_1,\, c_2$) with $p$ and $q$.} \label{tab:c1_c2}
	\vspace{.2cm}
	\centering
	\begin{tabular}{c|ccc|ccc}
		\hline
		Choice & $p$ & $c_1$ & $c_2$ & $q\, (\hat \xi=-.4)$ & $q\, (\hat \xi=-.25)$ & $q\, (\hat \xi=-.1)$ \\
\hline
{choice-1}& 6 & 10 & 5  &10& 8.5 & 7 \\
{choice-2}& 6 & 20& 7  & 13& 11& 8 \\
{choice-3}& 6 & 30& 9  &15& 13.5&9 \\
{choice-4}& 2 & 10& 5  &6 &4.5& 3 \\
{choice-5}& 2& 20& 7  & 9&7&4 \\
{choice-6}& 2& 30& 9  &11& 9.5 &5 \\
\hline
\end{tabular}
\end{table} 

\begin{figure}[!htb]
	\centering 
	\includegraphics[width=11cm, height=8cm]{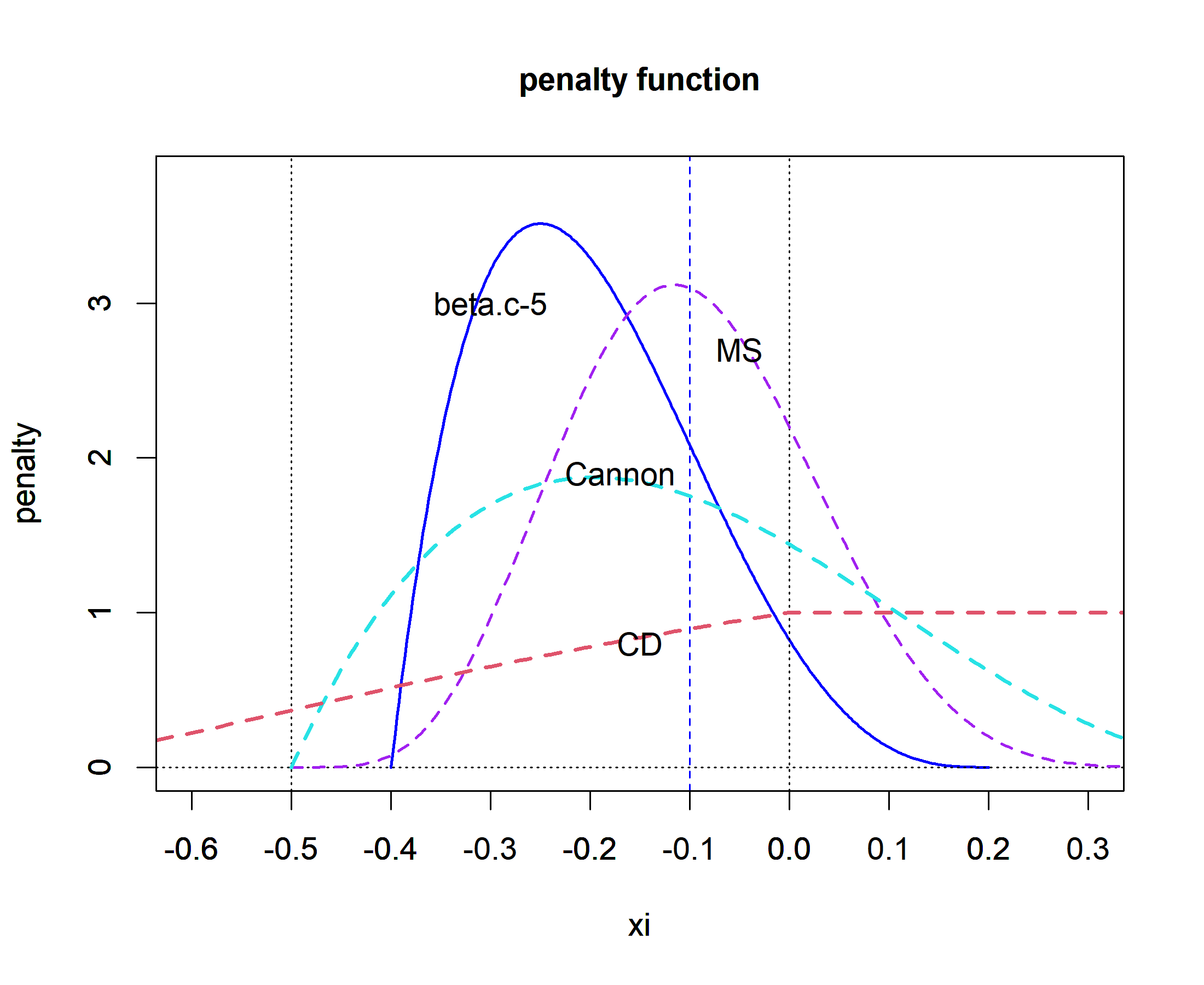} 
	\caption{Graphs of penalty functions (PFs) including Martins-Stedinger (MS), Coles-Dixon (CD), Cannon's PFs, and a beta PF with the choice-5 under the condition that $\hat \xi=-0.1$.}
	\label{fig:beta.comp}
\end{figure}

Figure~\ref{fig:pen.beta} shows graphs of these PFs for choices of hyperparameters and for different estimates of $\xi$ such as $\hat \xi = -0.4,\, -0.25,\, -0.1$. The upper panel of Figure~\ref{fig:pen.beta} is for $p=6$, while the lower panel is for $p=2$, with three sets of ($c_1,\,c_2$) values for each panel. By comparing graphs for each $\hat \xi$, graphs in lower panel shows more skewness than those in upper panel. As $\hat \xi$ decreases, a more pronounced shfit to the left is observed.
We see that the choice-6 is strongest in the aspect of the effect of PF, whereas the choice-1 is weakest among the above six choices of hyperparameters.  
Figure~\ref{fig:beta.comp} shows a comparison of the beta PF with the MS, CD, and Cannon PFs, where the beta PF is drawn for $\hat \xi =-0.1$ with the choice-5.


\section{GLME for the nonstationary model}
\subsection{Nonstationary model using L-moments}
The stationarity of (time-series) data is defined as a distribution that stays the same for any subsample of the original sequences. However, natural phenomena scarcely satisfy this assumption \citep{aghakouchak2012extremes, parey2021extreme}. 
In such a case when the stationarity is not satisfied, a nonstationary (NS) model-fitting technique is required. Thus the modeling for extremes of NS sequences have recently received more consideration in the conditions of climatic change \citep[for example]{slater2021nonstationary, castro2022practical, javan2023projected, grego2025robust}. 

The parameters of NS GEV model can be specified as:  
\begin{eqnarray} 
\mu(X)& =& \mu_0 + \mu_1 X_1 +\cdots + \mu_k X_k \label{gen_mu2} \\
\sigma(X)& =& \text{exp}\, \left( \sigma_0 + \sigma_1 X_1 +\cdots + \sigma_k X_k \right), \label{gen_sigma2}
\end{eqnarray}
where $X$ indicates a covariate vector and $k$ denotes the number of covariates. The covariates  may include time and physical variables. An exponential function in $\sigma(X)$ ensures the positivity of the scale parameter.
The shape parameters with covariates ($\xi (X)$) are challenging to estimate precisely; therefore, modeling $\xi$ as a smooth function of a covariate is usually unrealistic \citep{coles2001introduction, katz2002statistics}. Thus, we set $\xi$ as a constant in this study. 

Whereas applying the MLE to an NS GEV model is straightforward, applying LME is not. Hence, the MLE has been mostly used to estimate parameters of NS extreme value models. However, MLE is overly affected by large values or outliers toward the end of a sequence, causing inferior performance in such cases \citep{bousquet2021extreme, prahadchai2023analysis, shin2025building}.
As a robust alternative to the MLE, several studies considered the estimation methods based on L-moments \citep{mudersbach2010nonstationary, strupczewski2001non, strupczewski2016comparison, ribereau2008estimating, gado2016site}.

Recently, \cite{shin2025building} proposed an algorithm to build an NS GEV model using L-moments. Their method is regarded as an iterative refinement of the methods by \cite{strupczewski2001non} and  \cite{gado2016site}. The algorithm by \cite{shin2025building} is presented below, which is modified in the next section for applicable to the NS GLME method. 

\begin{itemize}
\item{ Step 1.} Fit $Z_X = \mu_0 + \mu_1 X_1 +\cdots + \mu_k X_k $ to the data using a robust regression method, to acquire the estimates of regression coefficients: ($\hat \mu_0, \ \hat \mu_1, \ \cdots, \ \hat \mu_k $). 
\item{Step 2.} Obtain absolute pseudo-residuals using $\epsilon_X\ =\ | Z_X - \hat \mu(X)|$.
\item{Step 3.} Fit $\epsilon_X\ =\ \text{exp}\, \left( \sigma_0 + \sigma_1 X_1 +\cdots + \sigma_k X_k \right)$ using ordinary regression, to obtain the least squares estimates: ($\hat \sigma_0, \hat \sigma_1, \cdots, \hat \sigma_k$).
\item{Step 4.} Determine $\mu_0, \; \sigma_0$ and $\xi$ which satisfy the following system of three equations, where $(\hat \mu_1, \cdots, \hat \mu_k)$ and $( \hat \sigma_1, \cdots, \hat\sigma_k)$ are fixed:
\begin{eqnarray} \label{gen-LME-NS}
\lambda_{G1}\ =\ l_1 (\tilde Z_X), ~~~
\lambda_{G2}\ =\ l_2 (\tilde Z_X), ~~~
\lambda_{G3}\ =\ l_3 (\tilde Z_X), 
\end{eqnarray} 
where $\lambda_{G1}, \; \lambda_{G2} $, and $\lambda_{G3}$ are the population L-moments of a standard Gumbel distribution, and $l_1(\tilde Z_X) , \; l_2(\tilde Z_X)$, and $l_3(\tilde Z_X)$ are sample L-moments computed from the transformed data $\{\tilde Z_X\}$. 
\end{itemize}

In Step 1, \cite{shin2025building} employed a robust regression using the function `lmrob' in `robustbase' package \citep{koller2017nonsingular} in R, which is the MM-estimator derived by \cite{yohai1987high}. 
In Step 4, the transformed data $\{\tilde Z_X\}$ are defined as follows:
\begin{equation} \label{transform}
\tilde Z_X = {{-1} \over {\hat\xi}} \;
\text{log}\ \left\{ 1 - \hat\xi \;\frac{Z_X -\hat\mu(X)}{\hat\sigma(X)} \right\}.
\end{equation}
Then, $\tilde Z_X$ follows the standard Gumbel distribution when the parameter estimates are true \citep{coles2001introduction}. \cite{shin2025building} found that this algorithm performs well 
 but sometimes underestimates the return level when $\xi \le -.2$. Thus, the bias correction (or GLME) method considered in the previous section for the stationary model would be also useful for this NS model approach. 

\subsection{GLME for the nonstationary GEV model}
For the NS GLME, we defined the NS GLD as follows:
\begin{equation} \label{NS-GLMDist}
{\widetilde {\omega} } \ (\mu_0, \, \sigma_0, \, \xi)\ = \ ( \underline{\bf \lambda_G} -\underline{\bf \tilde{l}}\ )^t \ {\bf \widetilde{V}}^{-1}\ (\underline{\bf \lambda_G} -\underline{\bf \tilde l }\ ) , 
\end{equation}
where $ \underline{\bf \lambda_G}=(\, \lambda_{G1}, \; \lambda_{G2} , ~ \lambda_{G3}\, )$, $\ \underline{\bf \tilde l} =(\, l_1(\tilde Z_X) , \; l_2(\tilde Z_X), \; l_3(\tilde Z_X)\, )$, and 
$\widetilde V$ is the $3 \times 3$ variance-covariance matrix of the sample L-moments of $\{\tilde Z_X\}$, up to the third order.

This study changes the Step 4 in the previous section  to the following {Step 4$^\prime$}, to make the NS GLD useful for cooperating with the PF:

\begin{itemize}
	\item {Step 4$^\prime$.} Determine $\mu_0, \ \sigma_0$ and $\xi$ that minimize the following function under the condition that $(\hat \mu_1, \cdots, \hat \mu_k)$ and $( \hat \sigma_1, \cdots, \hat\sigma_k)$ are fixed:
\begin{equation} \label{GLME-NS}
- \ln \ \left[ \ f_{NS}\ \{ \ {\widetilde{\omega} }\ ( \mu_0, \, \sigma_0, \, \xi ) \}\ \right] \; -\ \alpha_n \; \ln\ \{ p(\mu_0, \, \sigma_0, \, \xi) \}, 
\end{equation}
where 
\begin{equation} \label{MNorm-NS}
f_{NS}\ \{ \ {\widetilde {\omega} }\ ( \mu_0, \, \sigma_0, \, \xi) \}\ =\ \frac{1}{(2 \pi)^{3/2}\ |{\widetilde V}| } \;
\exp \left\{-{1\over2}\ {\widetilde {\omega} }\ ( \mu_0, \, \sigma_0, \, \xi) \right\} .
\end{equation}
\end{itemize}

The same PFs for the stationary model are employed for the NS model in this study, which means that $ p(\mu_0, \, \sigma_0, \, \xi)$ in (\ref{GLME-NS}) is actually $p(\xi)$ in (\ref{pen.norm}) or in (\ref{pen.beta}).

\section{Monte Carlo simulation}
\subsection{Simulation setting}
This study conducted a Monte Carlo simulation study to evaluate the estimation performance of the GLME. Random samples were generated from the GEVD. We fixed $\mu=100$ and $\sigma=30$ for the stationary case. The shape parameter was varied from $-.45$ to $.45$. The case $\xi < -0.5$ was not considered because a GEVD does not have a finite variance when $\xi < -0.5$. The focal point of interest in this work is the estimation of the 100-year return levels (0.99 quantile). We set the sample sizes as $n=30, \, 50, \, 70$ for the stationary case and $n=40, \, 70$ for the NS case. 

For a NS case in this study, we considered the following time-dependent NS GEV model with parameters:
\begin{eqnarray} \label{GEV11}
\mu(t)\ =\ \mu_0 + \mu_1 \times t , ~~~~
\sigma(t)\ =\ \text{exp}\ (\sigma_0 + \sigma_1 \times t),  ~~~~\xi(t)\ =\ \xi,
\end{eqnarray}
where we set $t = \text{year} - t_0 +1$, so that $t = 1, 2, \dots, n$. In this case, $t_0$ is the starting year of the observations and $n$ denotes the sample size (number of blocks). This model is referred to the GEV11. For its parameters, we set $\mu_0 = 0, \; \mu_1 =-.1, \; \sigma_0=1$, and $\sigma_1=.02$, following \cite{gado2016site}. This parameter setting generates a NS sample with decreasing location and increasing scale parameters as time evolves. As same as the stationary case, $\xi$ is varied from $-.45$ to $.45$.

To compare several estimators, the following evaluation measures were calculated:
\begin{equation} \label{eval}
\text{Bias} = \bar{\hat{r}} - r , ~~~~
\text{SE} = \left\{ {\frac{1}{N} \sum_{i=1}^N (\hat{r}_i -\bar{\hat{r}})^2} \right\}^{1/2}, ~~~~
\text{RMSE} = \left\{ { \frac{1}{N} \sum_{i=1}^N ({\hat{r}}_i - r)^2 }\right\}^{1/2} , 
\end{equation}
where $\hat{r}_i$ is the estimate of T-year return level for $i$-th simulation sample,  $r$ denotes the true return level which is already known, SE stands for the standard error of the estimator, RMSE indicates the root mean squared error, and $\bar{\hat{r}} =  \sum_{i=1}^N \hat{r}_{i} /N$, where $N$ indicates the number of trials ($N=1, 000$). Smaller values of the above measures are better than larger ones. 

For the NS case, the 100-year conventional (or `effective') return level \citep{gilleland2016extremes, katz2012statistical} at the end of the sample was considered in this study, although some `redefined' return levels for NS model are available \citep{olsen1998risk, parey2021extreme}. 
 The conventional $T$-year return level for the NS GEV model is obtained from
\begin{equation} \label{conv_rt}
r_T (t)= \mu(t)-\frac{\sigma(t)}{\xi}\Big[1-y_T^{-\xi}\Big],\  \text{for} \ \xi \neq 0,
\end{equation}
\noindent where $y_T =-\log(1-1/T)$. 

This simulation study examined only two (normal and beta) PFs to correct the bias of the LME in the GEV models, using the choice-3 for normal PF and choice-1 for beta PFs. For real data analysis, one can of course choose different PFs or priors based on the element of interest or prior information. 

\subsection{Simulation results}
\begin{figure}[!htb]
\centering
\includegraphics[width=16.5cm, height=7cm]{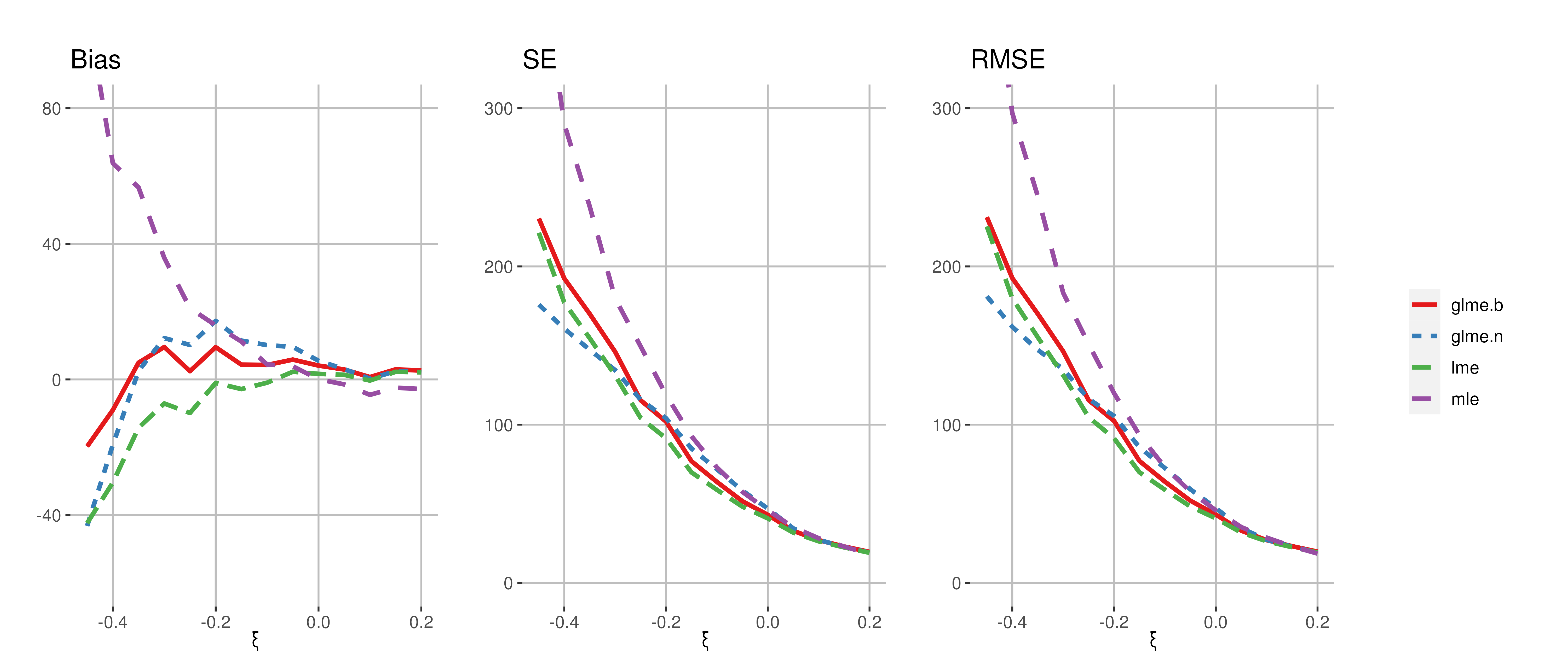} 
\caption{Plots of the bias, standard error (SE), and root mean squared error (RMSE) as $\xi$ changes for the stationary generalized extreme value  model for four estimation methods: maximum likelihood estimation (mle), L-moment estimation (lme), generalized LME with a normal penalty function with the choice-3 (glme.n), and GLME with a beta penalty function with the choice-1 (glme.b), drawn from 1,000 random trials for a sample size  $n=30$.} \label{fig:stnary_30_50}
\end{figure}

Figure~\ref{fig:stnary_30_50} illustrates the simulation results for the bias, SE, and RMSE of the stationary GEVD.  In this case, the 100-year return levels were computed using the four estimation methods with a sample size $n=30$, as $\xi$ varies from $-.45$ to .2. Tables of comparable results for sample sizes of $n=50$ and $n=70$ are presented in the Supplementary Material. Figure~\ref{fig:stnary_30_50} indicates that some negative bias exists in the LME and positive bias exists in the MLE, as $\xi$ approaches to $-.45$. 
The GLME with a beta PF is less biased than the others as $\xi$ approaches $-.45$. However, the GLME with a beta PF causes slight increases in the SE and RMSE compared to those of the LME.

\begin{figure}[!htb]
\centering
\includegraphics[width=16.5cm, height=7cm]{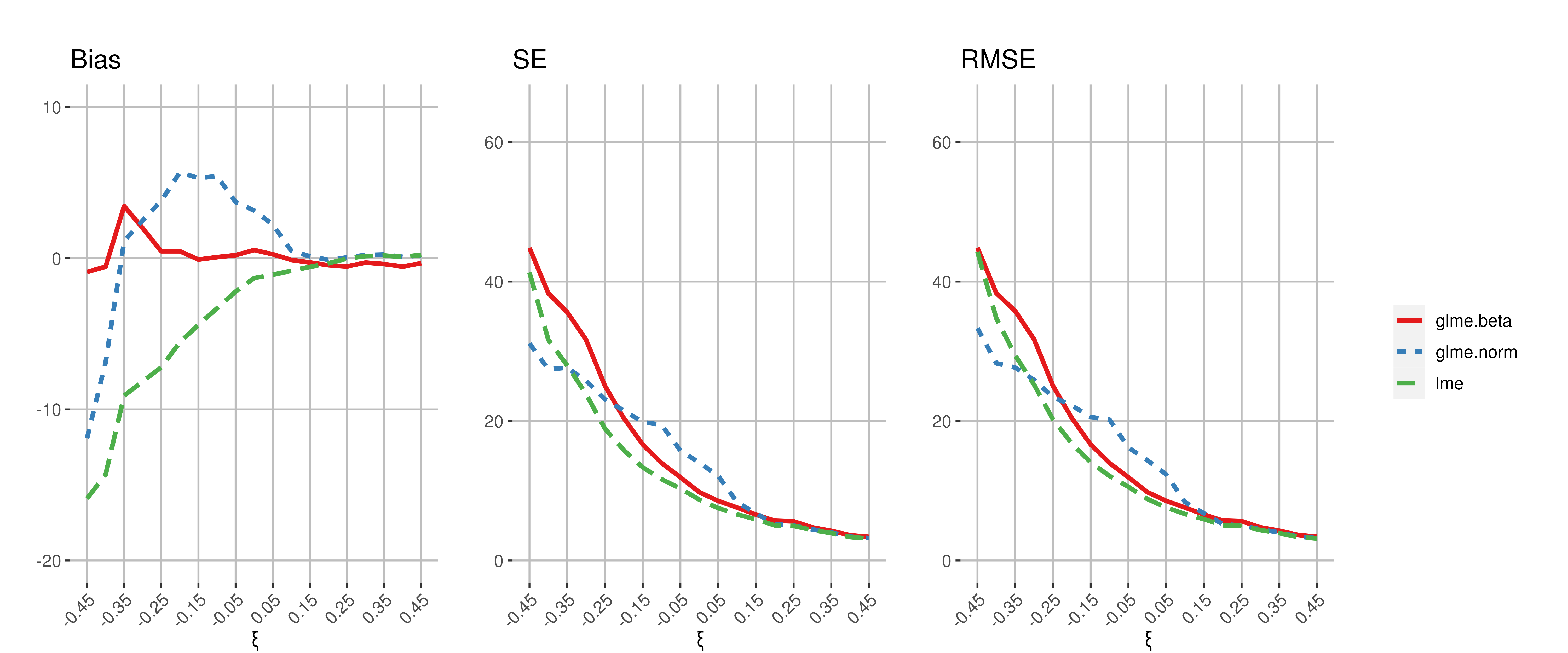} 
\caption{Same as Figure \ref{fig:stnary_30_50} but for the nonstationary GEV11 model  (\ref{GEV11}) and for a sample size $n=40$.} \label{fig:nonstnary_40_70}
\end{figure}

Figure~\ref{fig:nonstnary_40_70} illustrates the simulation results for the NS GEV11 model, which is the same as Figure~\ref{fig:stnary_30_50} but with a sample size $n=40$. Tables of similar results for a sample size of $n=70$ are offered in the Supplementary Material. Figure~\ref{fig:nonstnary_40_70} displays similar patterns as the stationary cases.

This simulation indicates that the negative biases of the LME for some $\xi$ are considerably corrected by the GLME with a beta PF with slight increases in the SE and RMSE of the LME for both stationary and NS GEV models, although the bias correction has room for improvement. 

\section{Real applications} \label{sec:realapp}
\subsection{Stationary model: US flood damage data}
For a real application, we considered the data for the total US economic damage (in billions of US Dollar) caused by floods for hydrologic years from 1932 to 1997. These data are available from the `extRemes' package \citep{gilleland2016extremes} of the R software. We analyze these data for the LOSSPW (loss per unit of wealth) variable with 66 observations (see \cite{katz2002statistics, pielke2000precipitation} for more information). The Mann--Kendall trend test found that $\tau = -.058$, and the two-sided $p$-value is 0.493. Therefore, we can treat these data as stationary.

\begin{table}[htb!]
\caption{Parameter estimates of the generalized extreme value distribution and estimated 50-year, 100-year, and 200-year return levels ($\hat{r}_{50},\,\hat{r}_{100},\,\hat{r}_{200}$) obtained using four methods for the US flood damage data (unit: billion USD). MLE: maximum likelihood estimation, LME: L-moment estimation, GLME.b.c-\# and GLME.n.c-\#: generalized LME with beta and normal penalty functions with hyperparameter choice-number, respectively. 
} \label{tab:losspw}
\vspace{.5cm}
\centering
\begin{tabular}{c|cccccccc}
\hline
Method & $\hat{\mu}$ & $\hat{\sigma}$ & $\hat{\xi}$ & $\hat{r}_{50}$ & $\hat{r}_{100}$ & $\hat{r}_{200}$ 
\\
\hline
MLE  & 119.17 & 102.09 & $-0.608$ & 1754 & 2709 & 4163 
\\
LME & 129.89 & 120.70 & $-0.377$ & 1205 & 1626 & 2172 
\\
\hline
GLME.n.c-1 & 128.89 & 120.03 & $-0.385$ & 1219 & 1651 & 2215 
\\ 
GLME.n.c-2 & 126.06 & 117.93 & $-0.405$ & 1248 & 1710 & 2320 
\\ 
GLME.n.c-3 & 128.32 & 119.63 & $-0.39$ & 1225 & 1664 & 2238 
\\ 
GLME.n.c-4 & 125.75 & 117.69 & $-0.407$ & 1251 & 1716 & 2330 
\\ 
\hline
GLME.b.c-1 & 125.44 & 117.43 & $-0.409$ & 1254 & 1721 & 2341 
\\ 
GLME.b.c-2 & 121.89 & 114.39 & $-0.429$ & 1277 & 1774 & 2441 
\\ 
GLME.b.c-3 & 119.77 & 112.44 & $-0.44$ & 1287 & 1798 & 2491 
\\ 
GLME.b.c-4 & 124.09 & 116.31 & $-0.417$ & 1264 & 1743 & 2381 
\\ 
GLME.b.c-5 & 119.65 & 112.33 & $-0.441$ & 1287 & 1800 & 2494 
\\ 
GLME.b.c-6 & 116.99 & 109.75 & $-0.453$ & 1295 & 1824 & 2546 
\\ 
\hline
\end{tabular}
\end{table}

Table~\ref{tab:losspw} lists the parameter estimates, the 50-, 100-, and 200-year return levels, and p-values obtained from 12 estimation methods. GLME.b.c-\# and GLME.n.c-\# represent the GLME with beta and normal penalty functions with hyperparameter choice-number, respectively. 
Note that the MLE of $\xi$ is larger negative value than $-0.6$ and so return levels from the MLE are very high compared to other estimates. All GLMEs of $\xi$ are more negative than that of the LME, which leads to larger return level estimates than that of LME. The cases of choice-1 for both normal and beta PFs provide the smallest return levels among those return levels obtained by GLMEs with each PF, although its are greater than the return levels obtained from LME. The choice-4 of normal PF and the choice-6 of beta PF provide the largest return levels for each PF.

\begin{figure}[!htb]
\centering
\includegraphics[width=12cm, height=11cm]{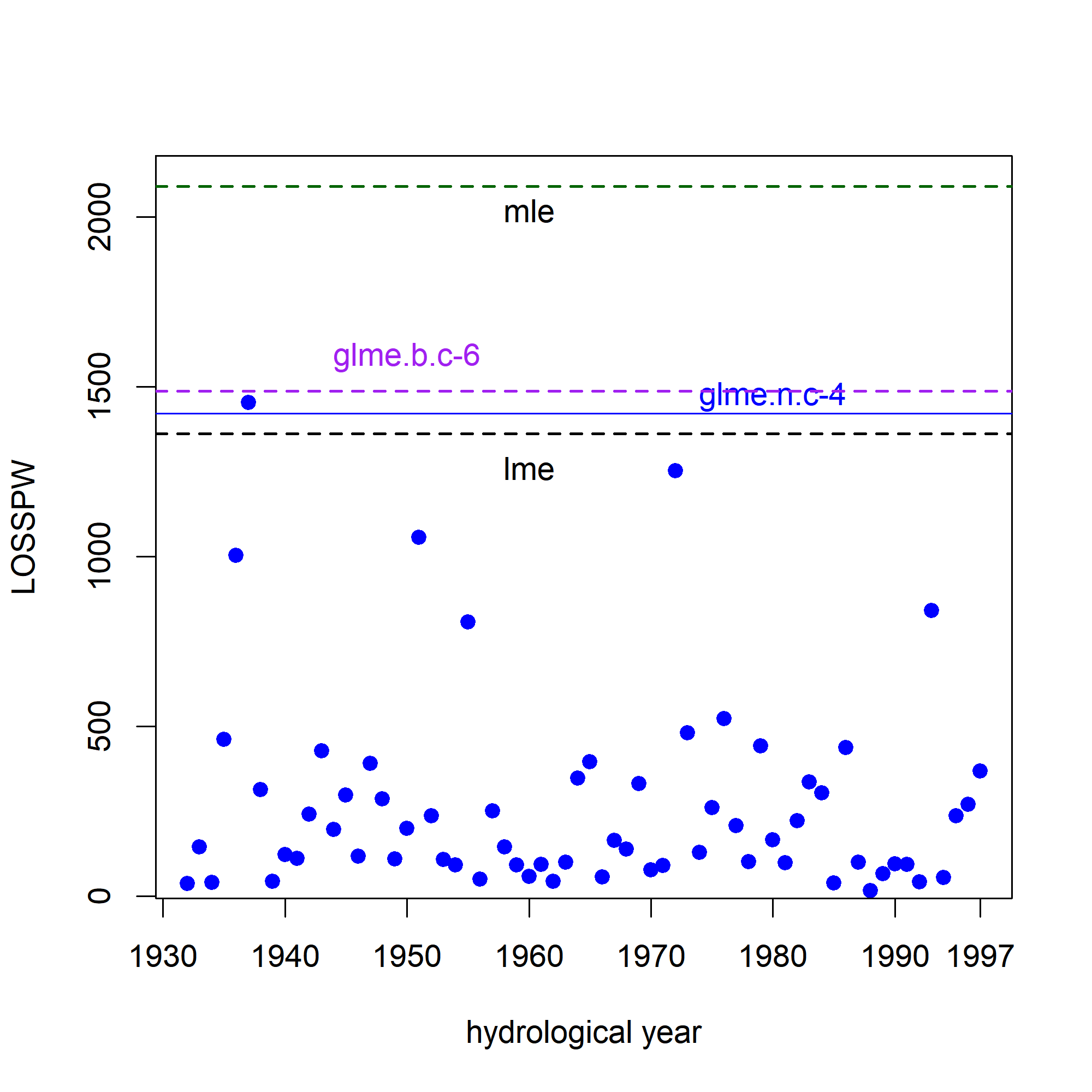} 
\caption{Same as Table~\ref{tab:losspw} but for scatterplot and 66-year return levels. 
} \label{fig:losspw}
\end{figure}

Figure~\ref{fig:losspw} presents the scatterplot of LOSSPW for the hydrological years and the 66-year return level estimates for the four methods (GLME.normal.choice-4, GLME.beta.choice-6, MLE, and LME). The reason why we consider the 66-year return level is because the sample size is 66, so that we can compare the largest observation with the 66-year return level.

	\begin{figure}[h!tb]
	\vspace{0.5 cm}
	\centering
	\begin{tabular}{l}
		\includegraphics[width=15cm, height=7cm]
		{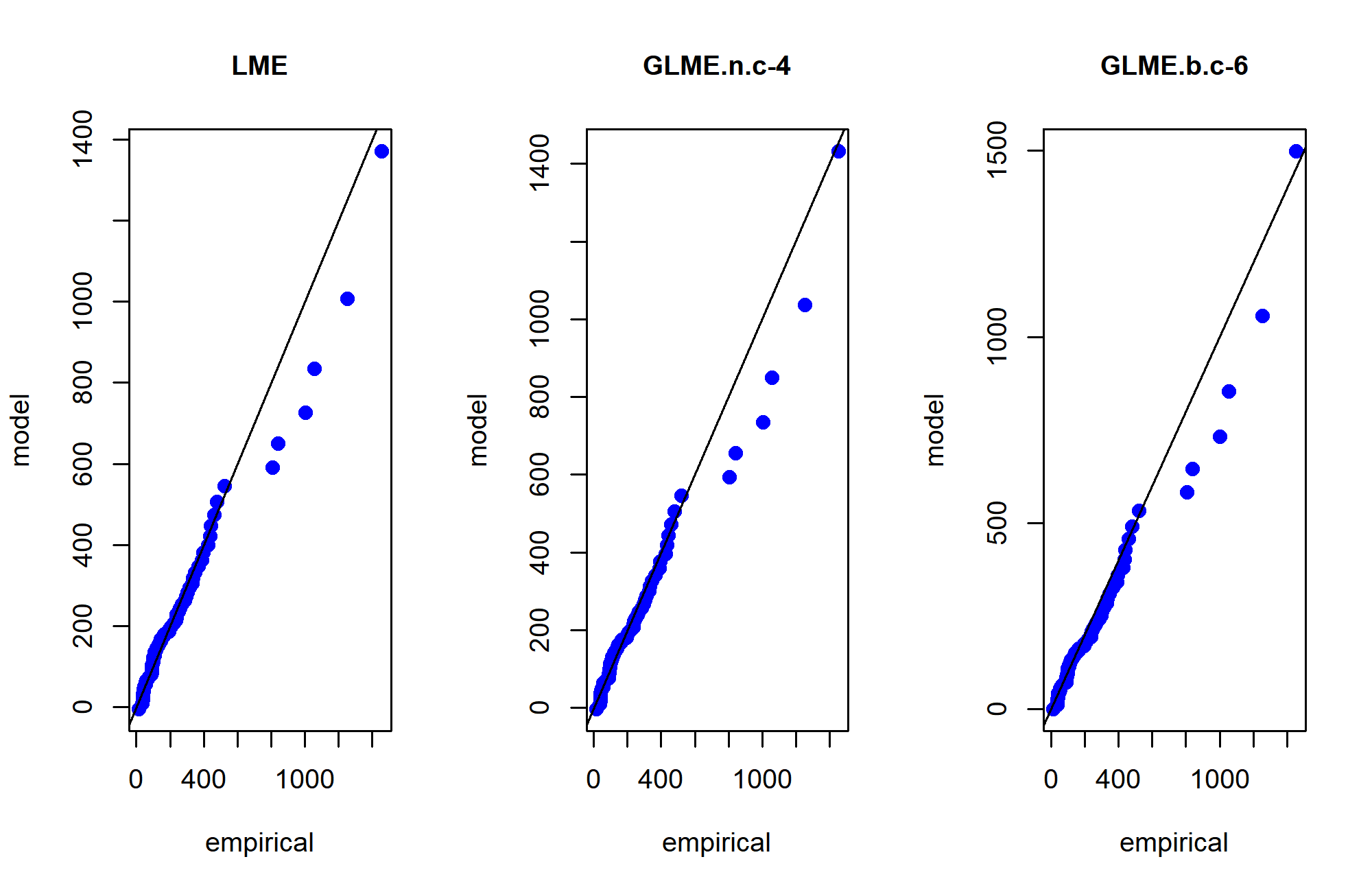} \end{tabular}
	\caption{Same as Table~\ref{tab:losspw} but for quantile-per-quantile plots from three LME methods. 
	} \label{fig:qq_losspw}
\end{figure}

	\begin{figure}[h!tb]
		\vspace{-0.3 cm}
		\centering
		\begin{tabular}{l}\includegraphics[width=12cm, height=9cm]{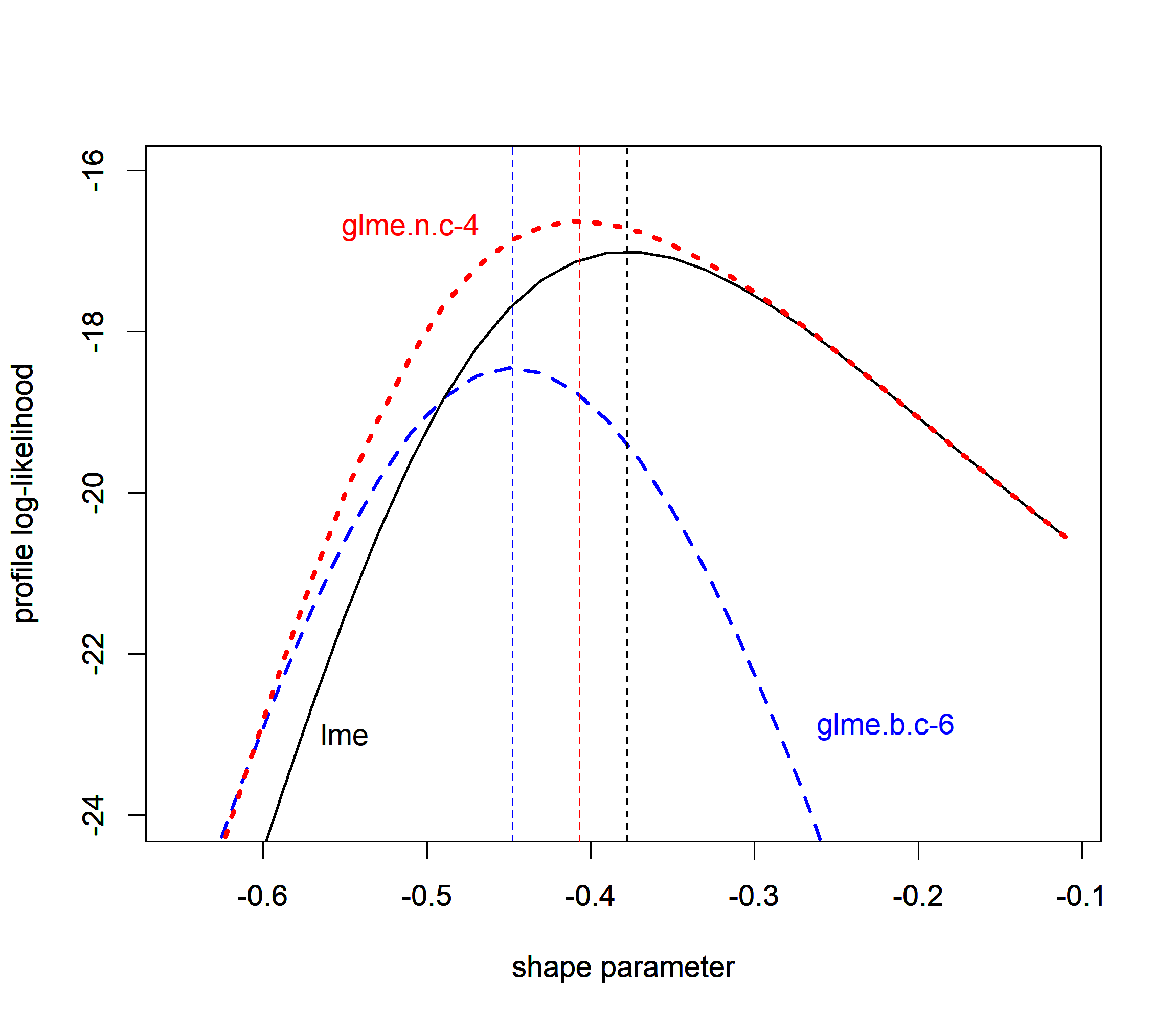} \end{tabular}
		\caption{Same as Table~\ref{tab:losspw} but for profile likelihood functions for $\xi$. 
		} \label{fig:proflike_losspw.xi}
	\end{figure}

Figure~\ref{fig:proflike_losspw.xi} illustrates the profile log-likelihood functions \citep{coles2001introduction} for $\xi$. The black line indicates the function of LME, whereas the red dotted and blue dashed lines indicate the functions of GLME.normal.choice-4 and GLME.beta.choice-6, respectively. The confidence interval obtained by GLME.beta.choice-6 is narrower than those by the other two. 
In summary, the GLME.b.c-6 seems to work better than others. 


\subsection{Nonstationary model: Maximum rainfall at Phliu Agromet}
For another application, we considered the annual maximum daily rainfall data for Phliu Agromet in Thailand. The dataset spans 40 years from 1984 to 2023, with units in millimeters (mm). The Thai Meteorological Department website (https://www.tmd.go.th) or  https://github.com/sygstat/GL-momentEst.git contains the data.

\begin{table}[!ht]
	\begin{center}
		\caption{Same as Table \ref{tab:losspw} but for the nonstationary GEV11 model fitted to annual maximum rainfall data (unit: mm) in Phliu Agromet in Thailand. 
		} \label{tab:phliu}
		\vspace{0.5cm}
		\begin{tabular}{c|ccccc|ccc} 
			\hline
			Method & $\hat \mu_0$ & $\hat\mu_1$ & $\hat\sigma_0$ & $\hat\sigma_1$ & $\hat\xi$ & $\hat{r}_{50}$ & $\hat{r}_{100}$ & $\hat{r}_{200}$ \\ \hline 
			MLE  & 118.62 & 1.092 & {3.05} & 0.022 & {$-0.072$} & {393} & 441 & 496\\ 
			LME & 121.26 & 0.936 & 2.95 & 0.028 & $-0.064$ & {423} & 478 & 535\\
			GLME.Cannon & 121.07& 0.936& 2.95&  0.028&  $-0.087$ & 436 &496 &560 \\ 
			\hline
            GLME.n.c-1 & 120.99 & 0.936 & 2.95 & 0.028 & $-0.098$ & 443 & 506 & 574 \\ 
            GLME.n.c-2 & 121.26 & 0.936 & 2.95 & 0.028 & $-0.064$ & 424 & 479 & 536 \\ 
            GLME.n.c-3 & 121.17 & 0.936 & 2.95 & 0.028 & $-0.075$ & 430 & 487 & 548 \\ 
            GLME.n.c-4 & 121.26 & 0.936 & 2.95 & 0.028 & $-0.064$ & 424 & 478 & 536 \\
			\hline
			GLME.b.c-1 & 121.17 & 0.936 & 2.95 & 0.028 & $-0.075$ & 430 & 487 & 548 \\ 
			GLME.b.c-2 & 121.08 & 0.936 & 2.95 & 0.028 & $-0.086$ & 436 & 496 & 560 \\ 
			GLME.b.c-3 & 120.99 & 0.936 & 2.95 & 0.028 & $-0.097$ & 442 & 506 & 573 \\ 
			GLME.b.c-4 & 121.08 & 0.936 & 2.95 & 0.028 & $-0.086$ & 436 & 496 & 560 \\ 
			GLME.b.c-5 & 120.9 & 0.936 & 2.95 & 0.028 & $-0.11$ & 450 & 517 & 589 \\ 
			GLME.b.c-6 & 120.72 & 0.936 & 2.96 & 0.028 & $-0.133$ & 466 & 540 & 621 \\
        \hline
		\end{tabular}
	\end{center}
	\vspace{0.1cm}
\end{table} 

\begin{figure}[!htb]
	\centering
	\begin{tabular}{l} \includegraphics[width=12cm, height=11cm]{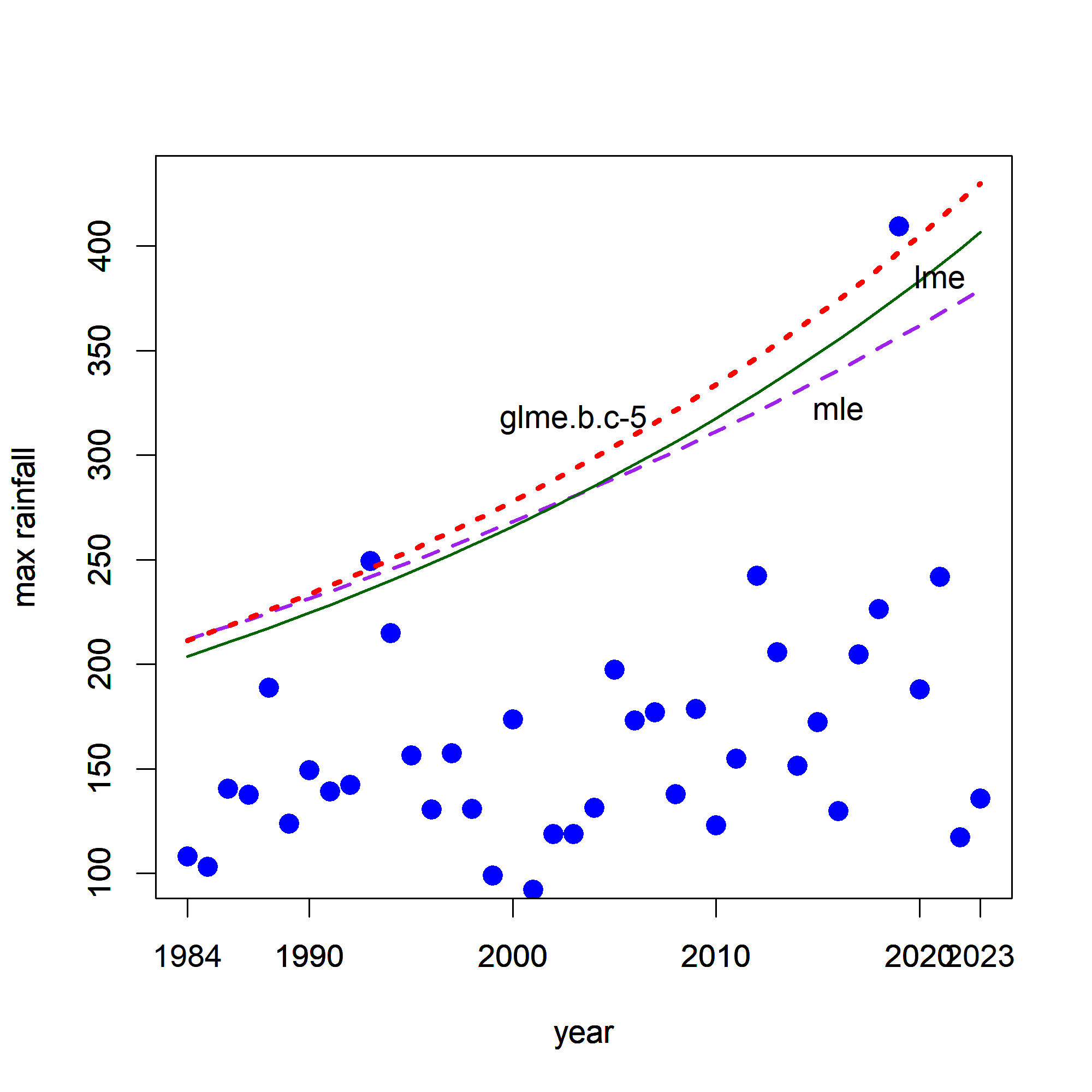} \end{tabular} 
	\caption{Same as Table \ref{tab:phliu} but for 40-year conventional return levels at every year. 
	}
	\label{rt-plot-phliu}
\end{figure}

Figure~\ref{rt-plot-phliu} displays a scatterplot of the data, displaying an increasing trend. To assess this trend, we conducted the Mann--Kendall trend test. The null hypothesis (no linear trend) was rejected at the 5\% level ($\tau=0.234,\, \text{p-value}=0.034$). Table~\ref{tab:phliu} lists the parameter estimates of GEV11 model, 50-year, 100-year, and 200-year conventional return level estimates at the end of sample. The GLMEs of $\xi$ are less than those from the MLE and LME applying the algorithm by \cite{shin2025building}, which results in larger return level estimates.

 Figure~\ref{rt-plot-phliu} also depicts the 40-year conventional return levels from the GEV11 model
 fitted by the three methods, including the MLE, the LME, and the GLME with beta PF choice-5. There are two observations above 40-year return level lines fitted from the LME and MLE. It seems the LME and MLE underestimate the return level around the end of sample. Whereas, it seems that GLME.b.c-5 provides 40-year return level comparable to two largest observations. The MLE may not work well in this case, because the sample size is relatively small compared to the number (five) of parameters to be estimated. 

\begin{figure}[h!tb]
	\vspace{-0.3 cm}
	\centering
	\begin{tabular}{l}\includegraphics[width=12cm, height=9cm]{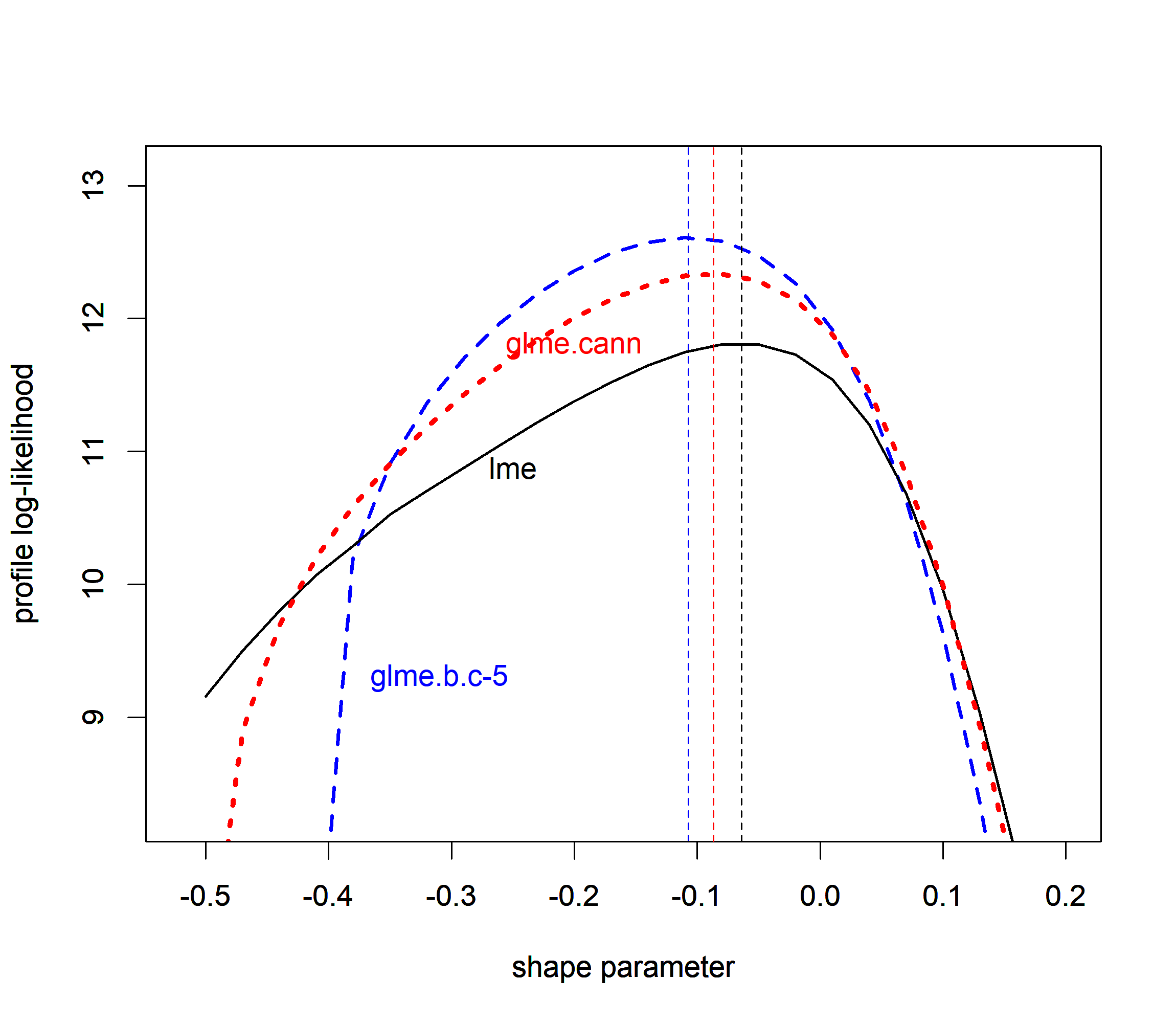} \end{tabular}
	\caption{Same as Table~\ref{tab:phliu} but for profile likelihood functions for $\xi$. 
	} \label{fig:proflike_phliu.xi}
\end{figure}

Figure~\ref{fig:proflike_phliu.xi} illustrates the profile log-likelihood functions \citep{coles2001introduction} for $\xi$. The black line indicates the function of LME, whereas the red dotted and blue dashed lines indicate the functions of GLME.Cannon and GLME.beta.choice-5, respectively. The confidence interval obtained by GLME.beta.choice-5 is narrower than those by the other two. 

In summary, these examples demonstrate that there are considerable benefits to be attained by using a GLME with data-adaptive beta PF 
in preference to the LME method.


\section{Discussion}

In constructing the generalized L-moment function in (\ref{GLME_func}), we set $\alpha_n = 1$. However, $\alpha_n$ may be selected from the data using the cross-validation technique such as employed by \cite{bucher2021penalized}. In addition, this study focused on only two (normal and beta) PFs on the shape parameter with a few choices of hyperparameters to correct the bias of the LME in the GEV models. Those PFs considered in this study are just examples used only for the bias correction of LME. For real applications, some other priors or different PFs, including the MS, CD, and Cannon's PF, can be allocated based on the element of interest or prior information. The data-adaptive normal PFs can be considered as well. In addition, one can assign PFs to the location and scale parameters too.


In this study, we obtained the mode of posterior as the GLME using an optimization technique. One can try to find the mean of posterior using Bayesian computation, which can lead to another GLME. The latter approach has pros and cons compared to the former approach. Anyway, the GLME with a properly chosen penalty function would provide better inferences in the aspect of utilizing additional information well, especially when sample size is small to moderate. 

\cite{shin2025modelaveraging_arXiv} considered a model averaging technique which employs the LME and Akaike information criterion to assign weights for candidate submodels. 
They concluded that the model averaging method based on L-moments performed well for heavy tail cases, but still showed a negative bias. We reasonably expect that their model averaging method may cooperate well to correct the bias with a penalty function or prior information and can be extended to Bayesian model averaging approach. One weakness of the proposed GLME method is that it takes more computing time than the LME. Bayesian model averaging may be much faster than the proposed GLME method. We defer this investigation to future research endeavor.

The proposed GLME can be applied similarly for the other estimation methods which employing, for example, so-called the ``LH-moments" \citep{wang1997lh, murshed2014lh, busababodhin2016lh} and generalized probability weighted moments \citep{ rasmussen2001generalized, ribereau2008estimating}.
In addition, the GLME method is applicable without further endeaver to an NS extreme value model with physical covariates in the parameters, though only a time-dependent NS model was illustrated in this study. 
Moreover, this work focuses on the GEVD only, but the proposed method can be applied to a broader range of models than the GEVD, including the generalized Pareto \citep{papastathopoulos2013extended, naveau2016modeling,
	kjeldsen2023use} and four-parameter kappa distributions \citep{hosking1994four, kjeldsen2017use, oshea2023improved, Strongetal2025kappa}. 

\section{Conclusion}
We proposed a generalized version of the LME method (GLME), which can cooperate with the penalty function (PF) or a prior for the parameters of the models, based on the generalized L-moments distance (GLD) and a multivariate normal approximation. 
We extended the GLME to nonstationary GEV model. The simulation study demonstrated that the generalized L-moment estimator can be less biased than the LME in the GEV models. In correcting the bias of LME using the GLME method, we suggested two (normal and beta) penalty functions including a data-adaptive one. It turned out that the data-adaptive beta PF seems more appealing than the normal PF. Applications to US flood damage data (for the stationary GEV model) and to maximum rainfall data at Phliu Agromet in Thailand (for the nonstationary GEV model) illustrate the usefulness of the proposed method.

This study may promote further work on penalized or Bayesian inferences based on L-moments, because the proposed GLME method cooperates with the PF or a prior on parameters which can be chosen based on the element of interest or prior information. 

\renewcommand{\baselinestretch}{0.8}
\begin{small}
	

\subsection*{Funding}
Jihong Park's research was supported by the BK21 FOUR (Fostering Outstanding Universities for Research, NO.5120200913674) funded by the Ministry of Education (MOE, Korea) and National Research Foundation of Korea (NRF). 
Yire Shin's research was supported by Basic Science Research Program through the
National Research Foundation of Korea (NRF) funded by the Ministry of Education (RS-2025-25436608).



\subsection*{Code and data availability}
The first version of the R code and the maximum rainfall data in Phliu Agromet in Thailand used in this study are available at https://github.com/sygstat/GL-momentEst.git.
The US flood damage data are available from the `extRemes' package~\citep{gilleland2016extremes} of the R software.

\subsection*{Supplementary Information}
This manuscript contains a pdf file for the Supplementary Material.

\subsection*{Conflict of interest}
The authors declare no potential conflicts of interest.

\subsection*{ORCID}
Yonggwan Shin: 0000-0001-6966-6511, ~ Yire Shin: 0000-0003-1297-5430,\\ ~ Jihong Park: 0009-0003-3191-9968, ~ Jeong-Soo Park: 0000-0002-8460-4869

\subsection*{Author Contributions}
 All authors contributed to the study design and conception, as well as to data collection, material preparation, computing, and analysis. The draft of manuscript was written by Yire and the last author, and the other authors provided comments on earlier versions. All authors then read and approved the final manuscript.

 	 \bibliographystyle{spbasic}
\bibliography{ref_glme_23Dec25}

\end{small}

\end{document}